\documentclass[11pt,a4paper]{article}
\usepackage[hyperref]{acl2020}
\usepackage{times}
\usepackage{latexsym}

\usepackage{booktabs}
\usepackage{color}
\usepackage{xcolor}
\usepackage{graphicx}

\newcommand\blfootnote[1]{%
  \begingroup
  \renewcommand\thefootnote{}\footnote{#1}%
  \addtocounter{footnote}{-1}%
  \endgroup
}

\usepackage{microtype}

\makeatletter
\DeclareRobustCommand{\EditStart}{%
  \@bsphack
  \leavevmode
  \color{blue}%
  \@esphack
}
\DeclareRobustCommand{\EditEnd}{%
  \@bsphack
  \normalcolor
  \@esphack
}

\aclfinalcopy 

\title{NARMADA: Need and Available Resource Managing Assistant for Disasters and Adversities}

\author{Kaustubh Hiware\textsuperscript{*} \\ Mercari, Inc\\Tokyo, Japan \\ hiwarekaustubh@gmail.com \\ \And 
    Ritam Dutt\textsuperscript{*} \\ Indian Institute of Technology\\
    Kharagpur, India\\
    ritam.dutt@gmail.com \\ \And
    Sayan Sinha \\ \hspace{5mm}Indian Institute of Technology\\ Kharagpur, India\\ sayan.sinha@iitkgp.ac.in \\
    \AND
    Sohan Patro \\ Microsoft Corporation\\ Redmond, USA \\ sopatr@microsoft.com \\ \And
    Kripabandhu Ghosh \\ Tata Consultancy Services\\ Pune, India \\ kripa.ghosh@gmail.com \\ \And
    Saptarshi Ghosh \\ Indian Institute of Technology \\ Kharagpur, India \\ saptarshi@cse.iitkgp.ac.in\\
}

\date{}

\begin{document}
\maketitle

\begin{abstract}
Although a lot of research has been done on utilising Online Social Media during disasters, there exists no system for a specific task that is critical in a post-disaster scenario -- identifying resource-needs and resource-availabilities in the disaster-affected region, coupled with their subsequent matching. To this end, we present NARMADA, a semi-automated platform which leverages the crowd-sourced information from social media posts for assisting post-disaster relief coordination efforts. 
The system employs Natural Language Processing and Information Retrieval techniques for identifying resource-needs and resource-availabilities from microblogs, extracting resources from the posts, and also matching the needs to suitable availabilities. The system is thus capable of facilitating the judicious management of resources during post-disaster relief operations. 
\end{abstract}

\section{Introduction}
\blfootnote{*Equal Contribution}

In recent years, microblogging sites like Twitter and Weibo have played a pivotal role in gathering situational information during disasters or emergency scenarios such as earthquakes, epidemic outbreaks, floods, hurricanes, and so on~\cite{social-media-emergency-survey, Nazer-disaster-osm-survey,disaster-info-mgmt-survey}. Specifically, there are two types of information which are considered useful (or `actionable') by rescue workers for assisting post-disaster relief operations.\footnote{We discussed with relief workers from `Doctors For You' (\url{http://doctorsforyou.org/}) and SPADE (\url{http://www. spadeindia.org/}).} 
These include (i)~\textbf{Resource needs} that talk about the requirement of a specific resource (such as food, water, shelter) and (ii)~\textbf{Resource availabilities} that talk about the availability of a specific resource in the region. 
Some examples of tweets that inform about resource-needs and resource-availabilities, taken from a dataset of tweets related to the 2015 Nepal earthquake, are shown in Table~\ref{tab:need-available-examples}.
We refer to such tweets as  `needs' and `availabilities' henceforth.

The two major practical challenges faced in this regard include (i)~automated {\it identification} of need and availability posts from social media sites such as Twitter and 
(ii)~automated {\it matching} of the appropriate needs and availabilities. 
There have been prior works which have tried to address each of these challenges separately. 
However, to the best of our knowledge, there exists no system that integrates the two tasks of identifying needs and availabilities and their subsequent matching.

\begin{table}[tb]
\vspace{2mm}
 \footnotesize
 \begin{tabular}{|p{0.4\columnwidth} | p{0.5\columnwidth} |}
 \hline
  \textbf{Needs (excerpts)} & \textbf{Availabilities (excerpts)}\\
  \hline
  Mobile phones are not working, no electricity, no \textbf{water} in \#Thamel, \#Nepalquake  &   Please contact for drinking free service \textbf{water} specially for Earthquake Victim. Sanjay Limbu [mobile num]
\\ \hline
Over 1400 killed.  Many Trapped. \textbf{Medical Supplies} Requested.   & 20,000 RSS personnel with \textbf{medical supplies} and other help the first to reach earthquake damaged zones in \#Nepal
\\  \hline
Nepal earthquake: thousands in need of \textbf{shelter}  in country little able to cope [url]   &   can anyone we know pick the 2000 second hand \textbf{tents} from Sunauli and distribute it to the people in need in Nepal?  
\\ \hline
\end{tabular}
\caption{ Examples of tweets stating resource-needs and tweets stating corresponding matching resource-availabilities, from a dataset of tweets on 2015 Nepal earthquake. The common resources for each pair are shown in boldface (table reproduced from our prior work~\cite{dutt2019utilizing}).}
\label{tab:need-available-examples}
\vspace*{-4mm}
\end{table}

In this work, we present
NARMADA ({\bf N}eed and {\bf A}vailable {\bf R}esource {\bf M}anaging {\bf A}ssistant for {\bf D}isasters and {\bf A}dversities),  a unified platform for the coordination of relief efforts during disasters by managing the resources that are needed and/or available in the disaster-affected region. 
NARMADA is designed to be a \textbf{semi-automated system} to ensure supervision and accountability. 

In this paper, we describe the Natural Language Processing and Information Retrieval techniques used in NARMADA for the following tasks -- (i)~identifying resource-needs and resource-availabilities from microblogs, (ii)~extracting resource names and other critical information from the posts (e.g., where the resource is needed, the quantity that is needed/available), and (iii)~matching the needs to suitable availabilities.
The system can be accessed from \url{https://osm-dm-kgp.github.io/Narmada/}.
Although the system is currently applied over tweets only, NARMADA can also seamlessly integrate information from other sources, as well as enable users to add new information as they deem fit. We believe that the use of this system during a real-time disaster event will help in expediting relief operations. 

Our work makes the following contributions.

1) We leverage contextual word embeddings to develop supervised models for automated classification of tweets that inform about need or availability of a resource.

2) We automate the process of categorising the type of resource present in needs and availabilities into food, health, shelter or logistics. This helps us to identify covert information present in tweets. 

3) We deploy NARMADA that leverages NLP and IR techniques to identify resource needs and availabilities from microblogs, extract relevant information, and subsequently match needs to suitable availabilities. We believe that such a system would assist in post-disaster relief operations.

\section{Related Work}

There has been a lot of recent work on utilising Online Social Media (OSM) 
to facilitate post-disaster relief operations -- see~\cite{social-media-emergency-survey,Nazer-disaster-osm-survey,disaster-info-mgmt-survey} for some recent surveys on this topic.  
For instance, there have been works on classifying situational and non-situational information~\cite{rudra-cikm-disaster,Rudra-tweb-2018}, location inferencing from social media posts during disasters~\cite{geotext,lingad,geolocalise-2018, savitr-smerp18, kumar2019location}, early detection of rumours from social media posts~\cite{Monda-2018}, emergency information diffusion on social media during crises~\cite{cindy-2018-ipm}, event detection~\cite{event-detect-2018}, extraction of event-specific informative tweets during disaster~\cite{lay-ipm} and so on. 
Tweets specific to particular disasters have been studied in \cite{gautam2019multimodal}, along with their categorisation. 
Certain other works have focused on the classification of such tweets by determining the probability of them being re-shared in Twitter \cite{neppalli2019predicting}.
A comparison of various learning-based methods has also been recently conducted in~\cite{assery2019comparing}.
 

Automated retrieval of needs and availabilities have been attempted by employing regular expressions~\cite{purohitFM}, pattern-matching techniques~\cite{emterms-iscram}, language models~\cite{BasuASONAM17}, and neural IR methods such as word and character embeddings~\cite{BasuASONAM17,Khosla2017}. 
Likewise, there has been prior research on the automated matching of the needs and availabilities using tf-idf similarity~\cite{purohitFM} and our prior works~\cite{matching-www18-poster,dutt2019utilizing} that used word-embeddings for the task. 
However, no prior work has attempted end-to-end identification and matching of needs and availabilities, which we attempt in this work.


Some information systems have also been implemented for disaster situations such as AIDR~\cite{AIDR} and Ushahidi~\cite{Ushahidi} which employs crowd-sourced information using social media to assist disaster operations. 
To our knowledge, none of the existing systems have attempted the specific tasks in this work -- identification and matching of resource-needs and resource-availabilities. 




\section{Dataset}

We reuse the dataset made available  by our prior works~\cite{Khosla2017,matching-www18-poster, dutt2019utilizing} which comprises tweets posted during two disaster events i.e. 
(i)~the earthquake in Nepal in April, 2015 \footnote{\url{https://en.wikipedia.org/wiki/April_2015_Nepal_earthquake}}, and 
(ii)~the earthquake in central Italy in August, 2016 \footnote{\url{https://en.wikipedia.org/wiki/August_2016_Central_Italy_earthquake}}. Henceforth, we refer to the scenarios as Nepal-quake and Italy-quake. 

The tweets were collected using the Twitter Search API\footnote{\url{https://dev.twitter.com/rest/public/search}} with the queries `nepal quake' and `italy quake'. 
The dataset consists of only English tweets since it was observed that most tweets are posted in English to enable rapid communication between international agencies and the local population. 

Removing duplicates and near-duplicates yielded a  corpus of \textit{50,068} tweets for Nepal-quake and \textit{70,487} tweets for Italy-quake. 
However, the number of tweets that inform about needs and availabilities is very low -- there are $499$ and $1333$ need and availability tweets for the Nepal-quake dataset. Likewise, the Italy-quake had only $177$ needs and $233$ availabilities (see~\cite{dutt2019utilizing} for more details). 


\section{Methodology}

In this section, we describe the methodologies that are incorporated within the NARMADA system. 
The overarching goal of the system is to facilitate post-disaster relief coordination efforts using the vast information available on social media. To that end, it performs three essential tasks --
(i)~identifying needs and availabilities, 
(ii)~extracting actionable information from the need and availability tweets,  and 
(iii)~matching appropriate needs and availabilities. 
NARMADA is designed to execute each of the above three tasks in an automated fashion. We elaborate on the specific methodology involved for each of these sub-tasks in the ensuing subsection. However, prior to each of these tasks, we perform pre-processing on the tweet text as follows. 

\vspace{2mm}
\noindent \textbf{Pre-processing tweets:} We employed standard pre-processing techniques on the tweet text to remove URLs (but not email ids), mentions, characters like brackets, `RT', and other non-ASCII characters like \#, \&, ellipses and Unicode characters corresponding to emojis. We also segmented CamelCase words and joint alphanumeric terms like `Nepal2015' into distinct terms (`Nepal' and `2015'). However, we did {\it not} perform case-folding or stemming on the tweet-text to enable subsequent detection of proper nouns (explained below).

\subsection{Identifying needs and availabilities}


Identifying needs and availabilities is challenging since they account for only $\approx3.64\%$ and $\approx0.58\%$ of the entire Nepal-quake and Italy-quake datasets, respectively. 
Prior works have approached this problem as a retrieval task using a wide array of techniques such as  regular-expressions~\cite{purohitFM},  pattern-matching~\cite{emterms-iscram},  language models~\cite{BasuASONAM17}, and recently neural IR techniques such as word and character embeddings~\cite{BasuASONAM17,Khosla2017, Khosla2019}.

To enable the real-time deployment, a system needs to filter out tweets on an individual basis. To that end, we decided to adopt a supervised approach for classifying a tweet as `need', or as `availability' or as `others' (i.e., a three-class classification problem). 
We experimented with different neural architectures for both in-domain and cross-domain classification. 
In-domain classification implies that the model is trained and tested on tweets related to the same disaster event.
On the other hand, cross-domain classification involves training on tweets related to one event (say `Nepal-quake') and evaluating on tweets related to another event (`Italy-quake')~\cite{Khosla2019}. 

\vspace{1mm}
\noindent {\bf Baseline methods:}
Convolutional Neural Networks (CNN) have been found to work well in the classification of disaster-related tweets~\cite{caragea2016identifying,nguyen2017robust}.
Hence we use the CNN of~\cite{kim-2014-convolutional} as a baseline model.  
We operate on 300-dimensional word-embeddings and fix the feature maps to 100 dimensions. We implement convolutional filters with kernel-size 3, 4, and 5 respectively, with stride 1, and non-linear ReLU activation units. Finally, we apply max-pooling before passing it through a fully-connected layer and softmax with negative log-likelihood (NLL) loss.  
We experiment with randomly initialized embeddings as well as different kinds of pre-trained embeddings --
Glove\cite{pennington2014glove}\footnote{\url{https://nlp.stanford.edu/projects/glove/}}, word2vec~\cite{word2vec} \footnote{\url{https://code.google.com/p/word2vec/}}, fasttetxt embeddings~\cite{fasttext-2017}\footnote{\url{https://fasttext.cc/docs/en/english-vectors.html}}   and CrisisNLP embeddings \cite{imran2016lrec} trained on tweets posted during many disaster events. 

\vspace{1mm}
\noindent {\bf Proposed model:}
We propose to use a pre-trained BERT model~\cite{devlin2018bert} (bert-base-uncased) to represent a tweet as a $768$-dimensional embedding. We pass the represented tweet through a fully connected layer which classifies it into the aforementioned three categories. Using BERT pre-trained embeddings helps us in two ways. Firstly, the BERT model itself remains a part of the entire end-to-end system; hence it gets fine-tuned while training. Moreover, BERT uses multiple bidirectional self-attention modules which helps capture contextual information. 

\vspace{1mm}
\noindent {\bf In-domain classification:}
Table~\ref{tab:need-offer-classification} notes the performances of the various classification models in {\it in-domain settings}, averaged over both the classes {\it needs} and {\it availabilities}. 
For each of the two datasets, we consider 20\% (randomly sampled) of the labelled data as the test set,  70\% of the rest as the training set, and the rest 10\% was used as the validation set.
We report the Precision, Recall and F1-score on the test set as the evaluation measures. We consider F1-score as the primary score since it incorporates both Precision and Recall.
The proposed BERT model outperforms all other models in terms of F1-score.

We trained two versions of our proposed BERT model --
(i)~one version was trained to optimise the F1-score (our primary measure) on the validation set, and
(ii)~the second version was trained to optimise the {\it Recall} on the validation set.
We specifically tried one version to optimise Recall, since it is usually considered important to identify all needs and availabilities in a disaster situation. 
As seen in Table~\ref{tab:need-offer-classification}, both versions of the model achieved comparable performance on the Nepal-quake dataset (F1-scores of $0.823$ and $0.828$). 
But the version trained for optimizing Recall achieved substantially higher performance on the Italy-quake dataset where needs and availabilities are much sparser. This improved performance justifies our decision of focusing on improving Recall.

\begin{table}[tb]
    \centering
    \resizebox{7.8cm}{!}{
    \begin{tabular}{|l|c|c|c||c|c|c|}
    \hline
    & \multicolumn{3}{|c||}{{\bf Nepal-quake}} & \multicolumn{3}{|c|}{{\bf Italy-quake}}\\ \hline
    Methodology & Prec & Rec & F1 & Prec & Rec & F1 \\ \hline
    CNN + random & 0.803 & 0.612 & 0.681 &0.926&0.552 & 0.637\\
    CNN + Glove & 0.790 & 0.668 & 0.716&0.846&0.680&0.727\\
    CNN + Word2vec & 0.796& 0.660 & 0.712&0.847&0.644&0.709 \\
    CNN + Fasttext & 0.771& 0.628& 0.683&0.870&0.640&0.703 \\
    CNN + CrisisNLP & 0.767 & 0.634 & 0.682 & 0.734& 0.585& 0.635 \\
    \textbf{BERT} (proposed for F1) & 0.786 & 0.866 & 0.823 & 0.856   & 0.722 & 0.779 \\
    \textbf{BERT} (proposed for Rec) & 0.791& 0.872& {\bf 0.828} & 0.843  &0.810 & {\bf 0.826}  \\
    
    \hline
    \end{tabular}}
    \caption{Performance of the neural architectures for in-domain classification of tweets into three classes -- needs, availabilities, and others. Best F1-scores in boldface.}
    \label{tab:need-offer-classification}
\end{table}

\vspace{2mm}
\noindent {\bf Cross-domain classification:}
In a cross-domain setting, the model is trained on tweets of one event and then evaluated on tweets of the other event.
We compare the performance of the  BERT model against the best-supervised model (`Best-SM') of~\cite{Khosla2019}, which is a CNN classifier initialised with CrisisNLP embeddings.
Table~\ref{tab:cross-1} shows results when the models are trained on Italy-quake tweets and tested on Nepal-quake tweets.
Similarly, Table~\ref{tab:cross-2} shows the opposite setting, i.e., the models are trained on Nepal-quake tweets and tested on Italy-quake tweets. In both cases, we use the BERT model optimised for F1-score, as described above. 
Even for cross-domain performance, we see that the BERT model outperforms the CNN-based baseline of~\cite{Khosla2019}.



\begin{table}[tb]
    \centering
    \small
    \resizebox{7.7cm}{!}{
    \begin{tabular}{|l|l|l|l|}
    \hline
    \textbf{Method} & \textbf{P@100} & \textbf{R@100} & \textbf{F1@100} \\ \hline
         \multicolumn{4}{|c|}{{\bf Needs}} \\ \hline
    Best-SM~\cite{Khosla2019} & 0.443 & 0.044 & 0.080  \\
    \textbf{BERT} (proposed) & 0.320 & 0.066 & {\bf 0.110} \\ 
    \hline
    
    \multicolumn{4}{|c|}{{\bf Availabilities}} \\ \hline
    Best-SM~\cite{Khosla2019} & 0.533 & 0.019 & 0.037 \\
    \textbf{BERT} (proposed) & 0.500 & 0.038 &  \textbf{0.070} \\
    \hline
    \end{tabular}}
    \caption{Performance of the neural architectures when trained on Italy-quake and tested on Nepal-quake. Best F1-scores in boldface.}
    \label{tab:cross-1}
\end{table}

\begin{table}[tb]
    \centering
    \footnotesize
    \resizebox{7.7cm}{!}{
    \begin{tabular}{|l|l|l|l|}
    \hline
    \textbf{Method} & \textbf{P@100} & \textbf{R@100} & \textbf{F1@100} \\ \hline
    \multicolumn{4}{|c|}{{\bf Needs}} \\ \hline
    Best-SM~\cite{Khosla2019} & 0.198 & 0.056 & 0.087  \\
    \textbf{BERT} (proposed) & 0.32 &  0.184& \textbf{0.234} \\ 
    \hline
    \multicolumn{4}{|c|}{{\bf Availabilities}} \\ \hline
    Best-SM~\cite{Khosla2019} &0.216 & 0.046 & 0.076 \\
    \textbf{BERT} (proposed) & 0.28 & 0.121&  \textbf{0.168} \\
    \hline
    \end{tabular}}
    \caption{Performance of the neural architectures when trained on Nepal-quake and tested on Italy-quake. Best F1-scores in boldface.}
    \label{tab:cross-2}
\end{table}

\subsection{Extracting relevant fields from needs and availabilities} 

Prior discussions with relief workers helped us identify the following five fields that are deemed relevant in coordinating the relief efforts, namely:
(i)~resource -- which items are needed/available,
(ii)~quantity -- how much of each resource is needed/available,
(iii)~location -- where is the resource needed/available,
(iv)~source -- who needs the resource or who is offering, and 
(v)~contact -- how to contact the said source. 




We adapt the unsupervised methodology of our prior work~\cite{dutt2019utilizing} to extract the relevant fields from needs and availabilities. 
We sought to incorporate this technique due to the paucity of labelled instances which discourages a supervised machine learning approach (and because gathering many labelled instances is difficult in a disaster scenario). 
Moreover, the unsupervised approach was shown to be generalizable across several datasets~\cite{dutt2019utilizing}. 
We describe the adapted methodology in this section. 


\vspace{2mm}
\noindent \textbf{Unsupervised resource extraction:} 
We start by giving a brief description of the methodology in~\cite{dutt2019utilizing}. 
We perform dependency parsing on the text to obtain a Directed Acyclic Graph (DAG). We compile an initial list of head-words (\textit{send, need, donate}, etc.) which consists of the verbs in the query-set and the ROOT word of the DAG. We have identified specific characteristics of the child nodes of the headwords that enable us to label the node as a potential resource. 

For example, if a word $w$ is tagged as a NOUN and is the direct object of the `donates', $w$ can be expected to be a potential resource. We have also identified dependency rules, that increases the list of head-words to improve our recall. We thus obtain a list of potential resources after dependency parsing. We then verify these potential resources by checking for the semantic similarity of the extracted words with a pre-compiled list of resources commonly used during disasters. The resource list is obtained from several reputed sources like UNOCHA\footnote{\url{https://www.unocha.org/}}, UNHCR\footnote{\url{https://www.unhcr.org/}} and WHO\footnote{\url{https://www.who.int/}}. This pre-compiled list also enables us to categorise the resources into four classes namely {\it food} (bottled water, biscuits, rice), {\it health} (blood, medicine, latrines), {\it shelter} (tents, blankets, tarpaulins), and {\it logistics} (electricity, helicopters, cash).

\vspace{2mm}
\noindent {\bf Adapting the method to deal with covert tweets:} One of the limitations of the unsupervised methodology in~\cite{dutt2019utilizing} is the inability to glean relevant information from {\it covert tweets} where the resource needed/available is not mentioned explicitly. 
We illustrate instances of such covert tweets in Table~\ref{tab:covert-tweets}. Since the resource name is not explicitly stated in the tweet-text, the methodology in~\cite{dutt2019utilizing} cannot identify the resources for such tweets.

\begin{table}[tb]
\centering
\small
\begin{tabular}{|l|l|} \hline
 Tweet Text & Resource \\ \hline
 villagers in the remote community of   & food\\
 ghyangphedi fear \textit{hunger} and \textit{\#starvation} & \\ \hline
 earthquake victims \textit{sleeping outside} in nepal & shelter\\ \hline
 people are \textit{shivering in the cold} & shelter \\ \hline
 free calls to italy in the wake of earthquake & logistics\\ \hline
\end{tabular} 
\caption{Examples of covert tweets and the corresponding resource class assigned to the tweet by our BERT-based resource classifier. 
}
\label{tab:covert-tweets}
\end{table}

To circumvent this problem, we again use the pre-trained BERT model~\cite{devlin2018bert} to encode a tweet. We pass this representation through a linear layer and perform multi-label classification into the aforementioned four categories, i,e. food, health, shelter and logistics. We use multi-label classification since a particular tweet can mention multiple resources. This adaptation helps the methodology to correctly classify many of the covert tweets, as demonstrated in Table~\ref{tab:covert-tweets} (the second column shows the resource-class that is assigned by our methodology).

\begin{table}[tb]
    \centering
  \footnotesize
    \begin{tabular}{|l|c|c|c|} \hline
    
      Dataset  & Precision & Recall & F1-score  \\ \hline
      Nepal-quake &  0.838 &0.882 &0.843\\
      Italy-quake &   0.825& 0.858& 0.823 \\ \hline

    \end{tabular}
    \caption{Performance of the multi-label BERT-based resource classifier on in-domain classification. }
    \label{tab:BERT-resource-classifier}
\end{table}

We report the {\it in-domain} classification performance of our BERT-based resource classifier for the Nepal-quake and Italy-quake datasets in Table~\ref{tab:BERT-resource-classifier}.  We test on 20\% of the data (sampled randomly) and train on the remaining 70\% while using 10\% for validation. We optimise the model with the highest macro F1-score on the validation set.

Next, we compare the performance of the proposed BERT-based resource classifier with that of the unsupervised methodology of~\cite{dutt2019utilizing} (which we refer to as `USM').
To ensure a fair comparison, we perform this comparison in a {\it cross-domain} setting wherein we train the supervised model on one dataset (e.g., Nepal-quake)  and evaluate on another (e.g., Italy-quake). 
We present the results of this comparison in Table~\ref{tab resource-cross}.

\begin{table}[t]
    \centering
    \small
    \resizebox{7.7cm}{!}{
    \begin{tabular}{|l|l|l|l|}
    \hline
    \textbf{Method} & \textbf{P@100} & \textbf{R@100} & \textbf{F1@100} \\ \hline
    \multicolumn{4}{|c|}{{\bf Nepal-quake}} \\ \hline
    USM~\cite{dutt2019utilizing} & 0.623 & 0.833 & {\bf 0.685}  \\
    \textbf{BERT} (trained on Italy) & 0.484 & 0.670 &  0.522 \\ 
    \textbf{BERT} (trained on Italy + 5\% Nepal) &
    0.636 & 0.834& 0.680\\
    \hline
    
    \multicolumn{4}{|c|}{{\bf Italy-quake}} \\ \hline
    USM~\cite{dutt2019utilizing} & 0.487 & 0.595 & 0.516  \\
    \textbf{BERT} (trained on Nepal) & 0.798 & 0.862 &  \textbf{0.808} \\ 
   
    \hline
    \end{tabular}}
    \caption{Comparing the BERT-based resource classifier with the unsupervised methodology (USM) of~\cite{dutt2019utilizing} in cross-domain setting. Best F1-scores in boldface.}
    \label{tab resource-cross}
    \vspace{-5mm}
\end{table}

\begin{table*}[tb]
 \small
 \centering
 \begin{tabular}
 {|p{0.34\textwidth} |p{0.12\textwidth}|p{0.12\textwidth}|p{0.1\textwidth}|p{0.10\textwidth}|p{0.06\textwidth}|}
 \hline
  \textbf{Tweet text (excerpts)} &\textbf{Resource} & \textbf{Location} &\textbf{Quantity} &\textbf{Source} & \textbf{Contact}\\
  \hline
  Urgent need of  analgesic,antibiotics, betadiene, swabs in kathmandu!! Call for help 98XXX-XXXXX  \#earthquake \#Nepal \#KTM (N) & analgesic, antibiotics, betadiene, swabs  & kathmandu, ktm, nepal && & 98XXX-XXXXX\\ \hline
  
  India sends 39 \#NDRF team, 2 dogs and 3 tonnes equipment to Nepal Army for rescue operations: Indian Embassy in \#Nepal (A) & NDRF team, dogs, & nepal & dogs - 2, NDRF team - 39 & India &\\ \hline 
  
   
   Visiting Sindhupalchok devastating earthquake highly affected district . Delivery Women in a tent . No water no toilet (N) & \textcolor{red}{tent, delivery women}, water &Sindhupalchok &  &&\\ \hline

    Rajasthan Seva Samiti donates more than 800 tents to Nepal Earthquake victims (A) & tents & & tents-800 & Rajasthan Seva Samiti &\\ \hline 
 \end{tabular}
 \caption{Examples of information extracted from need (N) and availability (A) tweets by the methodologies proposed in this work. Red colour indicates wrongly extracted information.}
\label{tab:field-extraction-examples}
\vspace*{-5mm}
\end{table*}

We observe from Table~\ref{tab resource-cross} that 
the BERT resource classifier trained on Nepal-quake significantly outperforms USM over the Italy-quake dataset (F1-score of $0.808$ for the BERT method and $0.516$ for USM). 
In contrast, the BERT resource classifier when trained on Italy-quake yielded significantly poorer results on Nepal-quake dataset than USM. 
However, training only on an additional 5\% of labelled instances of the Nepal-quake dataset, demonstrated comparative performance (F1-score of $0.680$ for the BERT method and $0.685$ for USM). 
The reason for these performances is as follows. 
The Italy-quake dataset does not contain mention of several amenities that are heavily prevalent in the Nepal-quake dataset, but {\it not} vice-versa. This difference is mainly because the Italy earthquake was a comparatively mild one in a developed region, and hence not many resources were needed; in contrast, the Nepal earthquake was a severe one in a developing region, and a lot of resources were needed in Nepal. Hence the Nepal-quake dataset contains mention of far more varied resources, as compared to the Italy-quake dataset.

Thus, including the BERT-based resource classifier in addition to the unsupervised methodology improves resource extraction performance, and also lends generalisability across different datasets.

\vspace{2mm}
\noindent \textbf{Extracting Locations:} We extract geographical locations from the tweet text using the methodology in our prior work~\cite{savitr-smerp18}. 
First, we apply several unsupervised techniques to extract a set of potential locations. These techniques include (i)~segmentating hashtags, (ii)~disambiguating proper nouns from parse trees, (iii)~identifying phrases with regex matches, (iv)~dependency parsing to locate nouns close from words in query-set in the DAG, and (v)~employing pre-trained Named Entity Recognizers \footnote{We use the inbuilt NER tool of SpaCy (\url{https://spacy.io/})} to identify words tagged as geographical location. 
Next, we verify these potential locations using a gazetteer. We consider those locations to be valid only if their geospatial coordinates lie within the boundary of the affected region (e.g., Nepal or Italy). We used two gazetteers namely Geonames \footnote{\url{http://www.geonames.org/}} and Open Street Map \footnote{\url{http:420//geocoder.readthedocs.io/providers/OpenStreetMap.html}} to identify locations with varying levels of granularity (as detailed in~\cite{savitr-smerp18}).  

\vspace{2mm}
\noindent \textbf{Extracting the source:} We consider as viable sources two types of words -- (i)~proper nouns that are tagged as organisations, persons or geographical locations by a Named Entity Recognizer, and (ii)~proper nouns that are child nodes of dependency parsing -- provided they have not been identified previously as `location' or `resources' during the verification phase. See our prior work~\cite{dutt2019utilizing} for details of the methodology.

\vspace{2mm}
\noindent \textbf{Extracting Quantity:} For each resource extracted, we identify whether it is preceded by a numeric token. The numeric token may be the orthographic notation of a number (e.g., `100') or may semantically represent a number (e.g., `hundred'). We assign the numeric token as the quantity of the particular resource. 

\vspace{2mm}
\noindent \textbf{Extracting Contact:} We use regular expressions to identify contacts corresponding to email-ids 
and phone numbers. 

\vspace{2mm}
\noindent The performance of our information extraction methods (in terms of precision, recall and F1-score) was similar to what is presented in~\citep{dutt2019utilizing}. In our experiments, we obtained F1-scores of $0.89$, $0.91$, $0.76$, $0.58$ and $1.00$ for identifying Resources, Location, Quantity, Source and Contact respectively, for need-tweets. Likewise, the F1-scores for availability-tweets were $0.85$, $0.85$, $0.84$, $0.65$ and $1.00$ respectively.
Table~\ref{tab:field-extraction-examples} shows some examples of the fields extracted by our methods from some need-tweets and availability-tweets.


\subsection{Matching needs and availabilities}

We propose a fast and real-time algorithm for matching needs and availabilities based on \textbf{proportion of common resources}.
Specifically, for a given need-tweet, we compute the match with a particular availability-tweet as the fraction of the resources extracted from the need-tweet, that are also present in the availability-tweet. 
For the given need tweet, availability-tweets are ranked in decreasing order of the fraction of common resources (ties resolved arbitrarily). 

We also experiment with some baseline methodologies, namely using common nouns~\cite{matching-www18-poster}, tf-idf vectors of the tweet text~\cite{purohitFM} and local word embeddings of the tweet~\cite{matching-www18-poster}. Our methodology (based on the proportion of common resources) obtains an F1-score of $0.84$ for Nepal-quake and an F1-score of $0.87$ for Italy-quake dataset respectively, which is competitive with the performance of the baselines. 


\vspace{3mm}
\noindent This section described the NLP and IR techniques used in NARMADA. The next section describes the system architecture.

\section{System Architecture}

\begin{figure}[tb]
    \centering
    \includegraphics[scale=0.4]{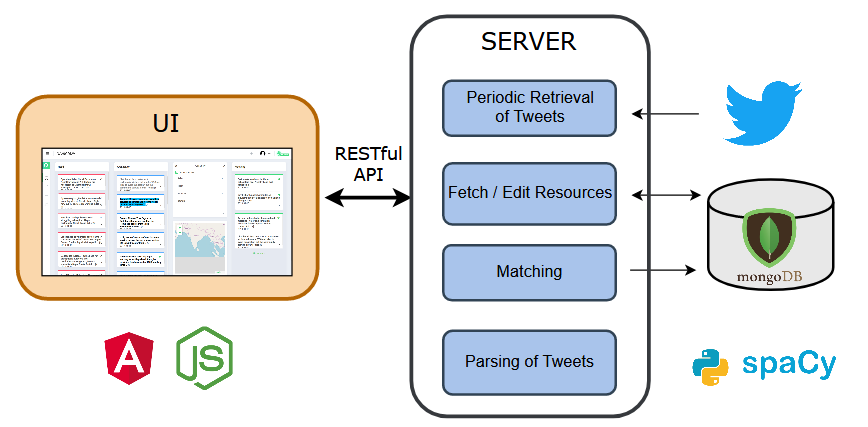}
    \caption{NARMADA's architecture overview}
    \label{fig:arch_overview}
\end{figure}
\vspace{-1mm}

The high-level system architecture for NARMADA is shown in Figure~\ref{fig:arch_overview}. 
The system can be accessed from \url{https://osm-dm-kgp.github.io/Narmada/}, where further details and a demonstration video are also provided\footnote{Additional details have been provided in the Appendices.}.
NARMADA is designed and built for the Web, thus not restricting it to any particular operating system or browser type, allowing cross-platform (desktop/mobile) functionality.

\begin{figure*}[htb]
    \centering

    \includegraphics[width=\textwidth]{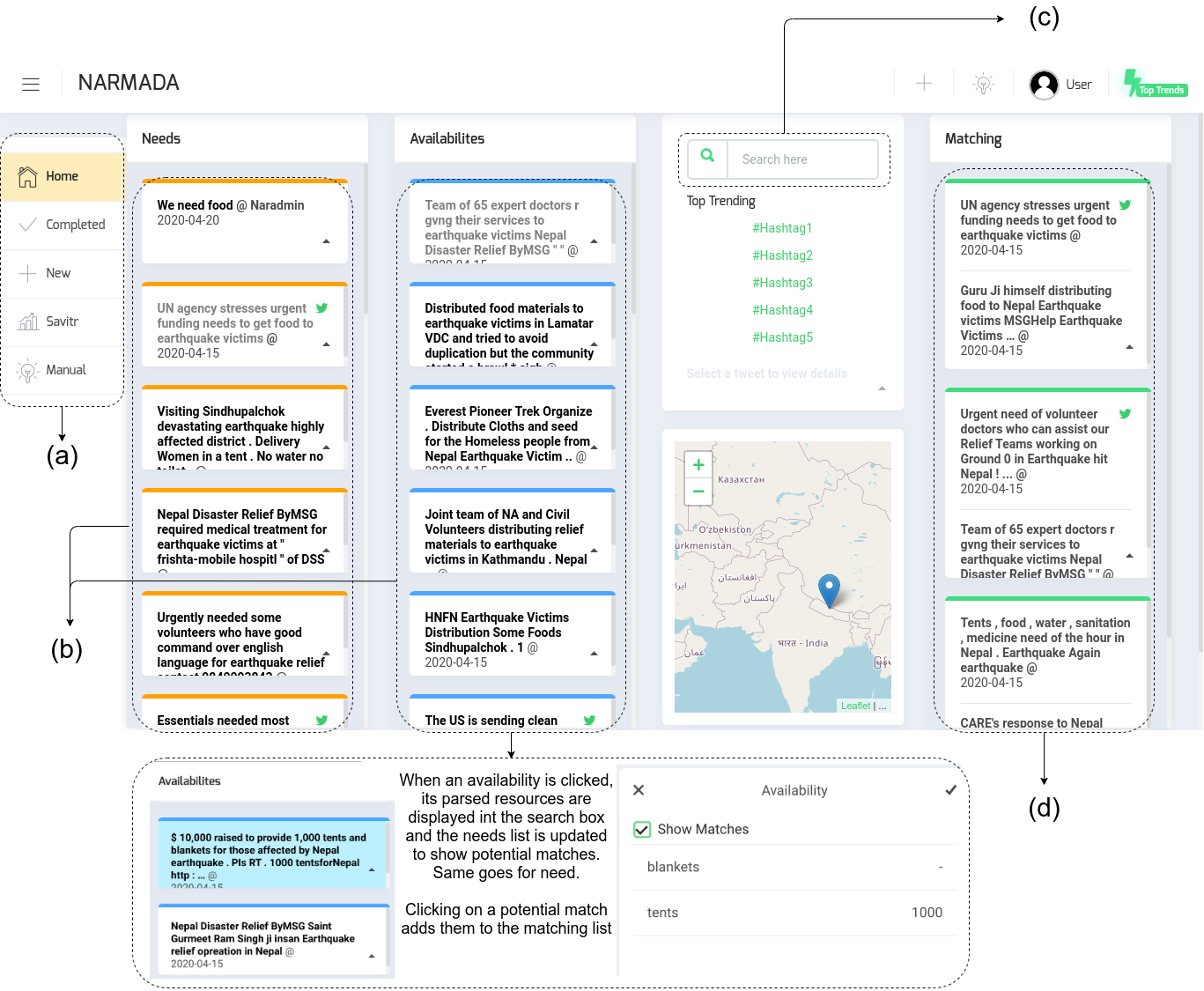}
    \caption{Dashboard of NARMADA -- (a)~\textbf{Navigation Buttons}. (b)~\textbf{Needs and Availabilities List}: tweets are displayed in reverse chronological order; gray tweet: already matched; black tweet: unmatched; each tweet contains a notch at the bottom-right corner, clicking on which reveals more details. (c)~\textbf{Search Box}: when a query is entered, the needs and availabilities containing the query-phrase are displayed. (d)~\textbf{Matching List}: displays the matched needs and availabilities; clicking a matching displays its resources, and gives the user an option to mark it as completed.}
    \label{fig:narmada_dashboard}
    \vspace{-3mm}
\end{figure*}




\subsection{User Interface}
\vspace{-1mm}

The user interface has been designed in Typescript using Angular, a popular web-application framework. ngx-admin\footnote{\url{https://github.com/akveo/ngx-admin}} was used as a boilerplate for front-end components. The interface has been designed to be intuitive, yet presenting as much information as possible without overcrowding. A detailed note is available at \url{https://osm-dm-kgp.github.io/Narmada/}.

The user interface comprises a dashboard (shown in Figure~\ref{fig:narmada_dashboard}) that acts as a landing page. Besides providing an initial view of active needs and availabilities (at the present point of time), it displays matched resources. The user is provided with various options to make it easy to search and locate resources as well as highlight items as deemed necessary.

An alternate section is available where users can enter new needs/availabilities manually. The class labels of the information are detected automatically, but the user is allowed to modify the same.  Another section for ``Completed matches" is to be used for logging completed or exhausted needs and resources. A user manual is also attached to the UI.


\vspace{-1.5mm}
\subsection{Server}
\vspace{-1.5mm}
The major services provided by the backend server include classification and categorisation of the tweets in the system. It also provides support for the addition of new information and their automatic categorisation. Facilities have been provided for marking resources once their need is fulfilled or the availability gets exhausted.

The server side uses NodeJS framework and is written in Javascript. Nginx is used as an HTTP server to make the frontend accessible to the public. However, the NLP-related extraction tasks are handled better in Python.
The server partly uses a Flask-based Python backend, a micro web framework.
The Flask server makes API calls to the deep learning classifiers, featuring BERT, which returns the output. The output is further reflected in the frontend. The server sends information requested by the user interface via \textit{RESTful API}, which supports cached responses on the frontend and enables the system to be scalable, thus allowing more users to use this service. 
API endpoints are publicly available, 
which would allow programmatic access to the server's functionalities (see \url{https://osm-dm-kgp.github.io/Narmada/}). 

\vspace{-1.5mm}

\section{Discussion}
NARMADA intends to assist in crossing the initial barrier in identifying and matching needs and availabilities from social media during the occurrence of a disaster. 
In practice, it becomes necessary for other service providers to be triggered in order to make sure that the needs are addressed, by proper collection, transportation and provisioning of the matched resources deemed to be available. 
For instance, the needs and availabilities could be marked on a map, with each type of resource being represented with a different symbol, making it easy to physically locate them. 
Local volunteers might be provided with a mobile app to help them find nearby needs and availabilities. Misinformation in twitter is common \cite{bal2020analysing}. The volunteers would also need a facility to confirm that the posted needs and availabilities are indeed genuine, concerning various parameters such as quantity (since at times of disasters, needs may be exaggerated). 


\vspace{-1.5mm}

\section{Conclusion and Future Work}
We proposed a system NARMADA for resource management during a disaster situation. 
Though the system is developed to work across posts from various social media platform, this research focused on data from Twitter. The real-time nature and easy access to large volumes of information provided by Twitter have made it a lucrative choice for disaster analytics. 

Currently, the system allows all users to perform any action on the system. One future task would be to implement a login system that would allow different access-levels to different users. 
For instance, a visitor would be able to only view and query information, a volunteer would be able to add new resources, mark a need as matched, etc., while a system administrator would have rights to undo all actions of all users, etc. 
The current system does not allow multiple volunteers to communicate within the platform over a resource, which we wish to incorporate in the future.
We also plan to incorporate support for vernacular languages, provided the requisite tools are available. 


\newpage
\bibliography{references}

\begin{thebibliography}{35}
\expandafter\ifx\csname natexlab\endcsname\relax\def\natexlab#1{#1}\fi

\bibitem[{Ush(2008)}]{Ushahidi}
 2008.
\newblock Ushahidi.
\newblock \url{https://www.ushahidi.com/}.

\bibitem[{AID(2015)}]{AIDR}
 2015.
\newblock Aidr (artificial intelligence for disaster response).
\newblock \url{http://aidr.qcri.org/}.

\bibitem[{Assery et~al.(2019)Assery, Xiaohong, Almalki, Kaushik, and
  Xiuli}]{assery2019comparing}
Nasser Assery, Yuan Xiaohong, Sultan Almalki, Roy Kaushik, and Qu~Xiuli. 2019.
\newblock Comparing learning-based methods for identifying disaster-related
  tweets.
\newblock In \emph{2019 18th IEEE International Conference On Machine Learning
  And Applications (ICMLA)}, pages 1829--1836. IEEE.

\bibitem[{Bal et~al.(2020)Bal, Sinha, Dutta, Joshi, Ghosh, and
  Dutt}]{bal2020analysing}
Rakesh Bal, Sayan Sinha, Swastika Dutta, Risabh Joshi, Sayan Ghosh, and Ritam
  Dutt. 2020.
\newblock Analysing the extent of misinformation in cancer related tweets.
\newblock \emph{arXiv preprint arXiv:2003.13657}.

\bibitem[{Basu et~al.(2017)Basu, Ghosh, Das, Dey, Bandyopadhyay, and
  Ghosh}]{BasuASONAM17}
Moumita Basu, Kripabandhu Ghosh, Somenath Das, Ratnadeep Dey, Somprakash
  Bandyopadhyay, and Saptarshi Ghosh. 2017.
\newblock Identifying post-disaster resource needs and availabilities from
  microblogs.
\newblock In \emph{Proc. ASONAM}.

\bibitem[{Basu et~al.(2018)Basu, Shandilya, Ghosh, and
  Ghosh}]{matching-www18-poster}
Moumita Basu, Anurag Shandilya, Kripabandhu Ghosh, and Saptarshi Ghosh. 2018.
\newblock Automatic matching of resource needs and availabilities in microblogs
  for post-disaster relief.
\newblock In \emph{Comp. Proc. WWW 2018 2018}, pages 25--26.

\bibitem[{Basu et~al.(2019)Basu, Shandilya, Khosla, Ghosh, and
  Ghosh}]{Khosla2019}
Moumita Basu, Anurag Shandilya, Prannay Khosla, Kripabandhu Ghosh, and
  Saptarshi Ghosh. 2019.
\newblock Extracting resource needs and availabilities from microblogs for
  aiding post-disaster relief operations.
\newblock \emph{IEEE Transactions on Computational Social Systems},
  6(3):604--618.

\bibitem[{Bojanowski et~al.(2017)Bojanowski, Grave, Joulin, and
  Mikolov}]{fasttext-2017}
Piotr Bojanowski, Edouard Grave, Armand Joulin, and Tomas Mikolov. 2017.
\newblock \href {https://doi.org/10.1162/tacl_a_00051} {Enriching word vectors
  with subword information}.
\newblock \emph{Transactions of the Association for Computational Linguistics},
  5:135--146.

\bibitem[{Caragea et~al.(2016)Caragea, Silvescu, and
  Tapia}]{caragea2016identifying}
Cornelia Caragea, Adrian Silvescu, and Andrea~H Tapia. 2016.
\newblock Identifying informative messages in disaster events using
  convolutional neural networks.
\newblock In \emph{International Conference on Information Systems for Crisis
  Response and Management}, pages 137--147.

\bibitem[{Devlin et~al.(2018)Devlin, Chang, Lee, and
  Toutanova}]{devlin2018bert}
Jacob Devlin, Ming-Wei Chang, Kenton Lee, and Kristina Toutanova. 2018.
\newblock Bert: Pre-training of deep bidirectional transformers for language
  understanding.
\newblock \emph{arXiv preprint arXiv:1810.04805}.

\bibitem[{Dutt et~al.(2019)Dutt, Basu, Ghosh, and Ghosh}]{dutt2019utilizing}
Ritam Dutt, Moumita Basu, Kripabandhu Ghosh, and Saptarshi Ghosh. 2019.
\newblock Utilizing microblogs for assisting post-disaster relief operations
  via matching resource needs and availabilities.
\newblock \emph{Information Processing \& Management}, 56(5):1680--1697.

\bibitem[{Dutt et~al.(2018)Dutt, Hiware, Ghosh, and Bhaskaran}]{savitr-smerp18}
Ritam Dutt, Kaustubh Hiware, Avijit Ghosh, and Rameshwar Bhaskaran. 2018.
\newblock Savitr: A system for real-time location extraction from microblogs
  during emergencies.
\newblock In \emph{Proc. WWW Workshop SMERP}.

\bibitem[{Gautam et~al.(2019)Gautam, Misra, Kumar, Misra, Aggarwal, and
  Shah}]{gautam2019multimodal}
Akash~Kumar Gautam, Luv Misra, Ajit Kumar, Kush Misra, Shashwat Aggarwal, and
  Rajiv~Ratn Shah. 2019.
\newblock Multimodal analysis of disaster tweets.
\newblock In \emph{2019 IEEE Fifth International Conference on Multimedia Big
  Data (BigMM)}, pages 94--103. IEEE.

\bibitem[{Hasan et~al.(2018)Hasan, Orgun, and Schwitter}]{event-detect-2018}
Mahmud Hasan, Mehmet~A. Orgun, and Rolf Schwitter. 2018.
\newblock {Real-time event detection from the Twitter data stream using the
  TwitterNews+ Framework}.
\newblock \emph{Information Processing \& Management}.
\newblock Online: https://doi.org/10.1016/j.ipm.2018.03.001.

\bibitem[{Imran et~al.(2015)Imran, Castillo, Diaz, and
  Vieweg}]{social-media-emergency-survey}
Muhammad Imran, Carlos Castillo, Fernando Diaz, and Sarah Vieweg. 2015.
\newblock {Processing Social Media Messages in Mass Emergency: A Survey}.
\newblock \emph{{ACM Computing Surveys}}, 47(4):67:1--67:38.

\bibitem[{Imran et~al.(2016)Imran, Mitra, and Castillo}]{imran2016lrec}
Muhammad Imran, Prasenjit Mitra, and Carlos Castillo. 2016.
\newblock Twitter as a lifeline: Human-annotated twitter corpora for nlp of
  crisis-related messages.
\newblock In \emph{Proceedings of the Tenth International Conference on
  Language Resources and Evaluation (LREC 2016)}, Paris, France. European
  Language Resources Association (ELRA).

\bibitem[{Karimzadeh et~al.(2013)Karimzadeh, Huang, Banerjee, Wallgr\"{u}n,
  Hardisty, Pezanowski, Mitra, and MacEachren}]{geotext}
Morteza Karimzadeh, Wenyi Huang, Siddhartha Banerjee, Jan~Oliver Wallgr\"{u}n,
  Frank Hardisty, Scott Pezanowski, Prasenjit Mitra, and Alan~M. MacEachren.
  2013.
\newblock Geotxt: A web api to leverage place references in text.
\newblock In \emph{Proceedings of the 7th Workshop on Geographic Information
  Retrieval}, pages 72--73.

\bibitem[{Khosla et~al.(2017)Khosla, Basu, Ghosh, and Ghosh}]{Khosla2017}
Prannay Khosla, Moumita Basu, Kripabandhu Ghosh, and Saptarshi Ghosh. 2017.
\newblock \href {http://arxiv.org/abs/1707.06112} {Microblog retrieval for
  post-disaster relief: Applying and comparing neural {IR} models}.
\newblock \emph{CoRR}, abs/1707.06112.

\bibitem[{Kim et~al.(2018)Kim, Bae, and Hastak}]{cindy-2018-ipm}
Jooho Kim, Juhee Bae, and Makarand Hastak. 2018.
\newblock {Emergency information diffusion on online social media during storm
  Cindy in U.S.}
\newblock \emph{{International Journal of Information Management}}, 40:153 --
  165.

\bibitem[{Kim(2014)}]{kim-2014-convolutional}
Yoon Kim. 2014.
\newblock \href {https://doi.org/10.3115/v1/D14-1181} {Convolutional neural
  networks for sentence classification}.
\newblock In \emph{Proceedings of the 2014 Conference on Empirical Methods in
  Natural Language Processing ({EMNLP})}, pages 1746--1751, Doha, Qatar.
  Association for Computational Linguistics.

\bibitem[{Kumar and Singh(2019)}]{kumar2019location}
Abhinav Kumar and Jyoti~Prakash Singh. 2019.
\newblock Location reference identification from tweets during emergencies: A
  deep learning approach.
\newblock \emph{International journal of disaster risk reduction}, 33:365--375.

\bibitem[{Laylavi et~al.(2017)Laylavi, Rajabifard, and Kalantari}]{lay-ipm}
Farhad Laylavi, Abbas Rajabifard, and Mohsen Kalantari. 2017.
\newblock {Event relatedness assessment of Twitter messages for emergency
  response}.
\newblock \emph{{Information Processing \& Management}}, 53:266--280.

\bibitem[{Li et~al.(2017)Li, Xie, Zeng, Zhou, Zheng, Jiang, Yang, Ha, Xue,
  Huang, Chen, Navlakha, and Iyengar}]{disaster-info-mgmt-survey}
Tao Li, Ning Xie, Chunqiu Zeng, Wubai Zhou, Li~Zheng, Yexi Jiang, Yimin Yang,
  Hsin-Yu Ha, Wei Xue, Yue Huang, Shu-Ching Chen, Jainendra Navlakha, and S.~S.
  Iyengar. 2017.
\newblock {Data-Driven Techniques in Disaster Information Management}.
\newblock \emph{ACM Comput. Surv.}, 50(1):1:1--1:45.

\bibitem[{Lingad et~al.(2013)Lingad, Karimi, and Yin}]{lingad}
John Lingad, Sarvnaz Karimi, and Jie Yin. 2013.
\newblock Location extraction from disaster-related microblogs.
\newblock In \emph{Proceedings of the 22nd international conference on world
  wide web}, pages 1017--1020. ACM.

\bibitem[{Mikolov et~al.(2013)Mikolov, Sutskever, Chen, Corrado, and
  Dean}]{word2vec}
Tomas Mikolov, Ilya Sutskever, Kai Chen, Greg~S Corrado, and Jeff Dean. 2013.
\newblock \href
  {http://papers.nips.cc/paper/5021-distributed-representations-of-words-and-phrases-and-their-compositionality.pdf}
  {Distributed representations of words and phrases and their
  compositionality}.
\newblock In C.~J.~C. Burges, L.~Bottou, M.~Welling, Z.~Ghahramani, and K.~Q.
  Weinberger, editors, \emph{Advances in Neural Information Processing Systems
  26}, pages 3111--3119. Curran Associates, Inc.

\bibitem[{Mondal et~al.(2018)Mondal, Pramanik, Bhattacharya, Boral, and
  Ghosh}]{Monda-2018}
Tamal Mondal, Prithviraj Pramanik, Indrajit Bhattacharya, Naiwrita Boral, and
  Saptarshi Ghosh. 2018.
\newblock {Analysis and Early Detection of Rumors in a Post Disaster Scenario}.
\newblock \emph{Information Systems Frontiers}, 20(5).

\bibitem[{Nazer et~al.(2017)Nazer, Xue, Ji, and
  Liu}]{Nazer-disaster-osm-survey}
Tahora~H. Nazer, Guoliang Xue, Yusheng Ji, and Huan Liu. 2017.
\newblock Intelligent disaster response via social media analysis a survey.
\newblock \emph{SIGKDD Explor. Newsl.}, 19(1):46--59.

\bibitem[{Neppalli et~al.(2019)Neppalli, Caragea, Caragea, Medeiros, Tapia, and
  Halse}]{neppalli2019predicting}
Venkata~Kishore Neppalli, Cornelia Caragea, Doina Caragea, Murilo~Cerqueira
  Medeiros, Andrea~H Tapia, and Shane~E Halse. 2019.
\newblock Predicting tweet retweetability during hurricane disasters.
\newblock In \emph{Emergency and Disaster Management: Concepts, Methodologies,
  Tools, and Applications}, pages 1277--1298. IGI Global.

\bibitem[{Nguyen et~al.(2017)Nguyen, Al-Mannai, Joty, Sajjad, Imran, and
  Mitra}]{nguyen2017robust}
Dat~Tien Nguyen, Kamla Al-Mannai, Shafiq~R Joty, Hassan Sajjad, Muhammad Imran,
  and Prasenjit Mitra. 2017.
\newblock Robust classification of crisis-related data on social networks using
  convolutional neural networks.
\newblock In \emph{ICWSM}, pages 632--635.

\bibitem[{Paule et~al.(2018)Paule, Sun, and Moshfeghi}]{geolocalise-2018}
Jorge David~Gonzalez Paule, Yeran Sun, and Yashar Moshfeghi. 2018.
\newblock {On fine-grained geolocalisation of tweets and real-time traffic
  incident detection}.
\newblock \emph{{Information Processing \& Management}}.

\bibitem[{Pennington et~al.(2014)Pennington, Socher, and
  Manning}]{pennington2014glove}
Jeffrey Pennington, Richard Socher, and Christopher~D. Manning. 2014.
\newblock Glove: Global vectors for word representation.
\newblock In \emph{Proc. EMNLP}.

\bibitem[{Purohit et~al.(2013)Purohit, Castillo, Diaz, Sheth, and
  Meier}]{purohitFM}
Hemant Purohit, Carlos Castillo, Fernando Diaz, Amit Sheth, and Patrick Meier.
  2013.
\newblock Emergency-relief coordination on social media: Automatically matching
  resource requests and offers.
\newblock \emph{First Monday}, 19(1).

\bibitem[{Rudra et~al.(2018)Rudra, Ganguly, Goyal, and Ghosh}]{Rudra-tweb-2018}
Koustav Rudra, Niloy Ganguly, Pawan Goyal, and Saptarshi Ghosh. 2018.
\newblock Extracting and summarizing situational information from the twitter
  social media during disasters.
\newblock \emph{ACM Trans. Web}, 12(3):17:1--17:35.

\bibitem[{Rudra et~al.(2015)Rudra, Ghosh, Goyal, Ganguly, and
  Ghosh}]{rudra-cikm-disaster}
Koustav Rudra, Subham Ghosh, Pawan Goyal, Niloy Ganguly, and Saptarshi Ghosh.
  2015.
\newblock Extracting situational information from microblogs during disaster
  events: A classification-summarization approach.
\newblock In \emph{{Proc. CIKM}}.

\bibitem[{Temnikova et~al.(2015)Temnikova, Castillo, and
  Vieweg}]{emterms-iscram}
Irina Temnikova, Carlos Castillo, and Sarah Vieweg. 2015.
\newblock {EMTerms 1.0: A Terminological Resource for Crisis Tweets}.
\newblock In \emph{{Proc. ISCRAM}}.

\end{thebibliography}
\bibliographystyle{acl_natbib}

\newpage
\newpage
\appendix
\onecolumn
\section*{Appendices}

\section{Detailed User Interface description}
Briefly, the user interface has five components:

\subsection{Dashboard}
As shown in Figure \ref{fig:narmada_dashboard}, the dashboard provides a preliminary view of unmatched needs and availabilities. Since the dashboard serves as the landing page, several functionalities are supported:

(a) It can view the \textbf{currently active needs, availabilities and matched resources} in separate tabs, as in Fig \ref{fig:narmada_dashboard}(b).
Additional details pertaining to a tweet like text, URL, \& parsed information like contact, location and source are displayed in a card layout.

(b) \textbf{Search} As seen in Fig \ref{fig:narmada_dashboard} (b),
when filled in, allows the viewers to query needs, availabilites and matched resources, all in one go.

(c) The interface displays the \textbf{matches corresponding to need and availabilities}. This allows one to view needs and matching availabilities (also vice versa) on the same screen. Clicking on a need resource reveals matching availabilities by default. The search tab is remodelled to show resource fields for the specific resource, along with an option to show potential matches
. Since we aim to build a semi-autonomous system, the sysadmin (or some volunteer) has to manually match/assign a need to an appropriate resource. Once a matching availability is selected, a match is made and appears in the Matching column.

(d) Once a match is made, assigned and completed by a volunteer, the match can be \textbf{marked completed}. Completed matches, needs and availabilities are explained in the next subsection.

(e) A map is provided that highlights the location the resource has been reported from, to assist in \textbf{geolocating resources}. 

\subsection{Completed matches}
This section acts as a logger to track completed or exhausted matches. It shares the same layout as the dashboard, apart from the presence of a search feature.

\begin{figure*}[htb]
    \centering
    \includegraphics[scale=0.3]{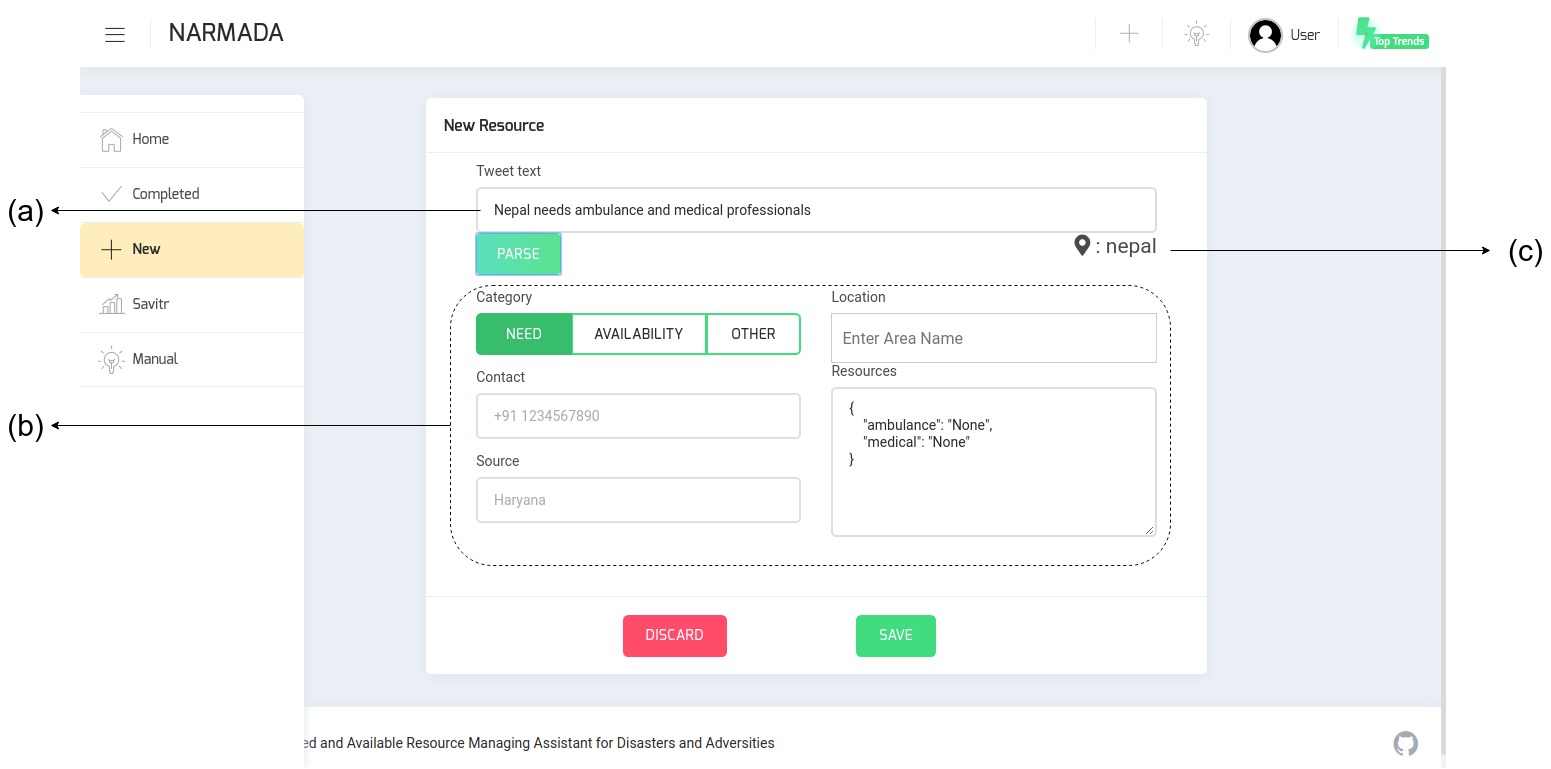}
    \caption{NARMADA's New Resource page. (a) The first step towards adding a new resource is to enter text and click on "PARSE". (b) All the parsed information are displayed, which can be modified by the user. (c) Auto-detected location from text.}
    \label{fig:narmada_new}
\end{figure*}

\subsection{New information}
As shown in Figure \ref{fig:narmada_new}, this view allows manual entry of details for a new need or availability. The information is automatically extracted but can be edited if required. The user is expected to enter text and click the parse button. A manual assignment must be made if classification could not be detected. Upon parsing (and editing) the information, the new resource can be saved, which is available on the dashboard immediately.

\subsection{Tweet traffic}
Apart from actionable items, users may be interested in the overall tweet activity about a particular disaster / word / phrase. For this, we integrate the view of Savitr \cite{savitr-smerp18}, which tracks Twitter activity related to any disasters on a day-to-day basis.

\subsection{Manual}
To allow users to view a sizeable amount of information at once, we understand it is possible to find the system complicated. Short, handy videos are available in order to explain the functionality for each of the prior components, along with a mission statement for the project.


\section{Detailed backend description}
The server side uses NodeJS framework and is written in Javascript. Nginx is used as an HTTP server to make the frontend accessible to the public. However, the NLP-related extraction tasks are handled better in Python. So a part of the server-side has been hosted with Flask, a micro web framework in Python. The Flask server makes API calls to the deep learning classifiers, featuring BERT, which returns the output. The output is further reflected in the frontend. The server sends information requested by the user interface via RESTful API, which supports cached responses on the frontend and enables the system to be scalable, thus allowing more users to use this service. 

API endpoints are publicly available, which would allow programmatic access to the server’s functionalities. API documentation can be found at \url{https://osm-dm-kgp.github.io/Narmada/#api-description}. The major services provided are:

\subsection{Fetching information i.e. needs, availabilites and matches}
Filtering by multiple conditions (such as matched or not, containing a particular resource) is also possible.

\subsection{Matching needs and availabilities}
For a provided need/availability, top 20 matches are suggested based on resource similarity.

\subsection{Elevating matched status}
Whenever a suitable match is found for a need/ availability, the corresponding pair is marked as \textit{Matched}. Once the Match has been assigned to a volunteer and is completed, the sysadmin can mark this match as \textit{Completed}, which moves both these resources from the dashboard to the Completed Resources view.

\subsection{Parsing and adding new information}
The system allows creation of new need/availability for a provided text. This is achieved by parsing all information - resources, contact, location, quantity and source from the said text and returning these fields.

\end{document}